\documentclass[aps,prb,twocolumn,superscriptaddress]{revtex4-1}
\usepackage{amssymb}
\usepackage{natbib}
\usepackage{epsf}
\usepackage{epsfig}
\usepackage{graphicx}
\usepackage{dcolumn}
\usepackage{braket}
\usepackage{bm}
\usepackage{amsfonts}
\usepackage{amsmath}
\usepackage{amssymb}
\usepackage{color,soul}
\usepackage{times}

\newcommand{\bea}{\begin{eqnarray}}
\newcommand{\eea}{\end{eqnarray}}

\makeatletter
\renewcommand{\p@subsection}{}
\renewcommand{\p@subsubsection}{}
\makeatother

\begin{document}

\title{On the Nature of Defects in Liquid-Phase Exfoliated Graphene}
\author{M. V. Bracamonte$^*$} 
\affiliation{Instituto de F\'{\i}sica Enrique Gaviola (CONICET) and FaMAF, Universidad Nacional de C\'ordoba, Argentina}
\author{G. I. Lacconi}
\affiliation{INFIQC, Departamento de Fisicoqu\'{\i}mica, Facultad de Ciencias Qu\'{\i}micas, Universidad Nacional de C\'ordoba, Argentina}
\author{S. Urreta}
\affiliation{Instituto de F\'{\i}sica Enrique Gaviola (CONICET) and FaMAF, Universidad Nacional de C\'ordoba, Argentina}
\author{L. E. F. Foa Torres}
\affiliation{Instituto de F\'{\i}sica Enrique Gaviola (CONICET) and FaMAF, Universidad Nacional de C\'ordoba, Argentina}

\vspace{0.5cm}

\begin{abstract}
Liquid-phase exfoliation is one of the most promising routes for large scale production of multilayer graphene dispersions. These dispersions, which may be used in coatings, composites or paints, are believed to contain disorder-free graphene multilayers. Here we address the nature of defects in such samples obtained by liquid-phase exfoliation of graphite powder in N-methyl--2--pyrrolidone. Our Raman spectroscopy data challenges the assumption that these multilayers are free of bulk defects, revealing that defect localization strongly depends on the sonication time. For short ultrasound times, defects are located mainly at the layer edges but they turn out to build up in the bulk for ultrasonic times above 2 h. This knowledge may help to devise better strategies to achieve high-quality graphene dispersions.

\end{abstract}

\date{\today} 
\maketitle

\section{Introduction} 
Graphene, the ultimate two-dimensional form of carbon, was discovered less than a decade ago.\cite{Novoselov2004,Novoselov2005,Zhang2005} Since that seminal discovery, graphene has united many of the electrical,\cite{Geim2007} thermal \cite{Balandin2008} and mechanical \cite{Lee2008} records known to man. Besides allowing the study of puzzling properties predicted for Dirac massless fermions \cite{Geim2007,Book}, graphene and the related materials also attract an unprecedented attention from technology \cite{Novoselov2012}. Indeed, its versatility allows for a wide range of expected applications including large displays \cite{Bae2010}, optoelectronic devices and ultracapacitors, and also conductive inks that may allow for ubiquitous printed electronics.\cite{Emtsev2009,Torrisi2012,Han2013}

Such a broad spectrum of applications requires very different production methods. On one hand, CVD stands as the most viable technique for large-area samples.\cite{Bae2010} On the other hand, mass-production of micrometer-sized samples for conductive inks requires cost-effective alternatives such as liquid-phase exfoliation of graphite.\cite{Nicolosi2013} By immersion of graphite powder in a suitable solvent, it can be exfoliated by using ultrasound (20-100 kHz). The process is simple and effective: ultrasonic waves produce a cavitation process which ultimately leads to the graphite exfoliation, while the solvent prevents the exfoliated multilayers from re-stacking. The simplicity and relatively low-cost of this method has triggered a lot of attention \cite{Nicolosi2013,Torrisi2012,Khan2011,Paton2014} and it has been demonstrated to be effective for the production of inkjet printed transistors.\cite{Torrisi2012} Unlike early methods involving oxidation of graphite followed by liquid exfoliation,\cite{Stankovich2006} here the $\pi$-orbitals are not disrupted and therefore the samples are expected to have higher conductivity and a reduced defect density. Concerning this latter technique, however, a few crucial questions remain open: What is the nature of the defects at the origin of the observed defect density evidenced by the D-band in the Raman spectra? Does the D-band stem only from the edges,\cite{Khan2010} or are there bulk defects? Is it possible to control the character of the dominant defects by changing the external parameters?

In this paper we shed light on some of these questions. We present results for liquid phase exfoliated graphene multilayers in N-methyl--2--pyrrolidone (NMP) obtained after different ultrasonic times and further characterized through Raman spectroscopy, UV--Vis spectrophotometry and scanning electron microscopy (SEM) techniques. From our statistical study of the correlation of the relative intensities of the D and G bands with the width of the latter we show the existence of a transition between samples with edge-dominated defects and bulk-dominated defects. This transition occurs as the ultrasonic time is increased indicating that actually shorter ultrasound times may help to obtain high-quality samples. Our analysis of the Raman spectra suggests that the defects are neither vacancies nor sp$^3$-like defects, leaving the formation of topological defects as a result of the cavitation process as the more likely alternative.

\section{Experimental Procedure}
Among the many different solvents that could be used to produce the graphene dispersions such as N,N--Dimethylacetamide (DMA), and
N,N--Dimethylformamide (DMF), we chose N--methyl--2--pyrrolidone (NMP). NMP is one of the best candidates because of its high boiling point and heat of vaporization, which help reducing the coffee-ring effect when drying after printing. Besides, NMP also improve the relative stability of the produced dispersions (we checked this in our samples, which remained stable for more than 6 months). Furthermore, it is known that the surface tension of this solvent is similar to that of carbon-based materials (40–50 mJ/m$^2$), thereby allowing graphite exfoliation.

The graphene dispersions reported in this paper were prepared by adding graphite powder (grade \#38, Fischer Scientific) to NMP at a concentration of $3.0$ mg mL$^{-1}$ and then exposing the dispersions to ultrasound (sonic bath Testlab, TB02, $40$kHz with a power of $80$W). The sonication times ranged between $30$ and $500$ minutes, while keeping water bath temperature below $32^{\circ}{\rm C}$. All samples were sonicated at the same spot in the sonic bath. After sonication, the dispersions were centrifuged (centrifuge Arcano, 80-2B) at 500 rpm for 30 min. After centrifugation the top 80\% of the supernatant was collected.

UV--Vis absorption spectroscopy was performed with a Shimazdu UV-1700 spectrophotometer using quartz cuvette with 1.0 cm optical path. RAMAN spectra were collected with a LABRAM-HR Horiba Jobin-Yvon confocal microscope at 514 nm and a 100x objective lens with a numerical aperture $0.9$. To avoid sample damage or laser-induced heating, the incident power was kept below 1 mW.  The RAMAN spectra were collected on numerous spots on the sample deposited onto 300 nm $SiO_2$/$Si$ wafer. The spectra have been deconvoluted with Lorentzian line shapes for all peaks (these line-shapes give the best $r^2=0.99$). The intensity ratio $i(D)/i(G)$ was obtained after a baseline correction. All measurements were performed at room temperature.

Scanning electron microscopy (SEM) images were obtained with a Field Emission Gun Scanning Electron Microscope (FE-SEM, Zeiss, $\Sigma$IGMA model) working at 10keV.

\section{Results and discussion}

\subsection{Effect of sonication time on graphene concentration}

\begin{figure}[tb]
\includegraphics[width=0.95\columnwidth]{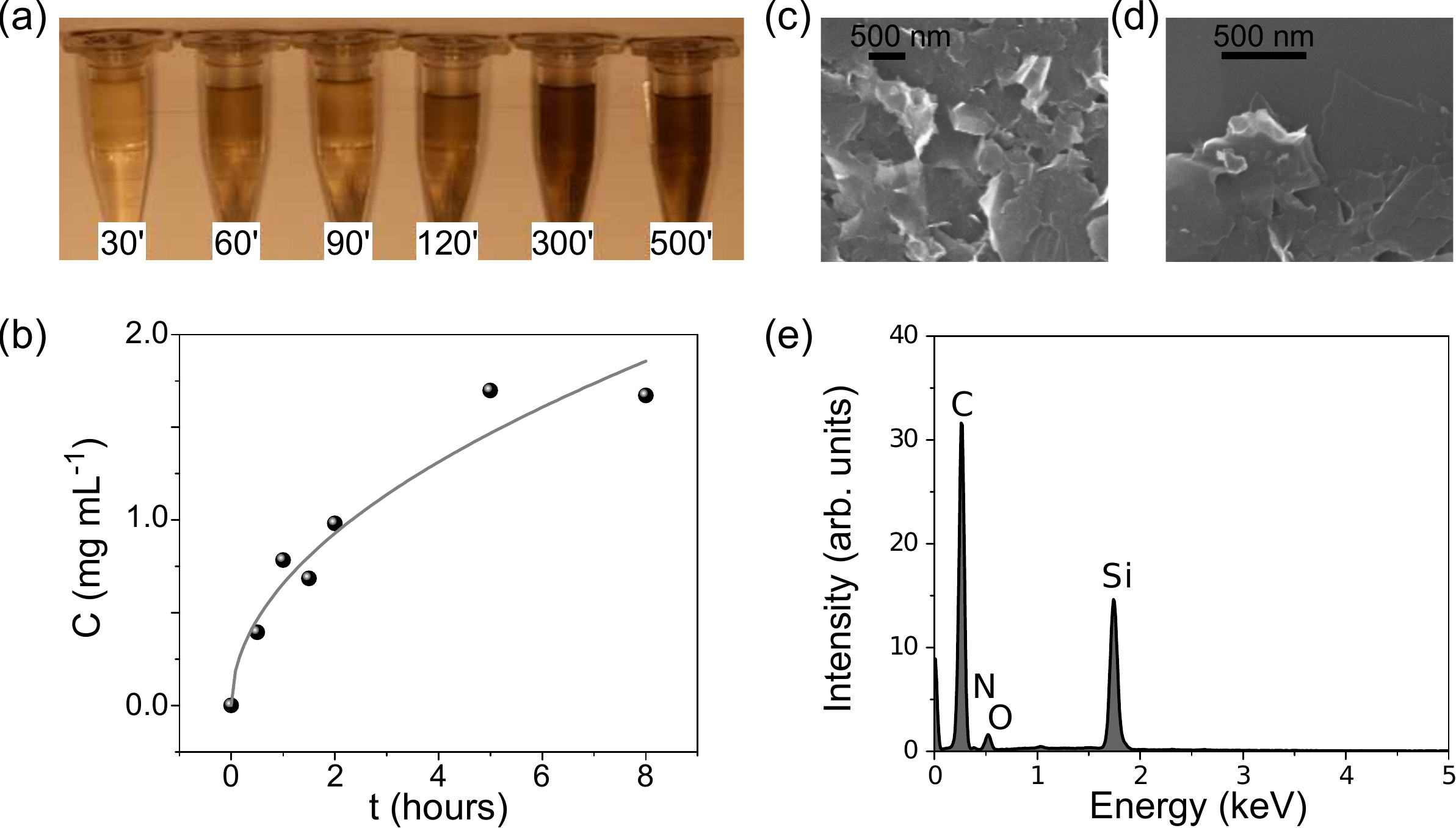}
\caption{(a) Dispersions of graphene multilayers in NMP. The darker the color the higher the concentration of the dispersion. (b) Concentration $C$ of dispersed graphene as a function of sonication time $t$. Dots represent the experimental data and the solid line the best fit obtained with a square root dependence $C = k t^{1/2}$. (c-d) Typical SEM images of flakes deposited on n-doped Si substrate (sonication time of 120 minutes). (e) EDS spectra for flakes sonicated for $t=300$ min confirming the absence of a significative ammount of oxides or other contaminants.}
\label{fig1}
\end{figure}

Graphene dispersions with a concentration (graphite powder in NMP) of 3.0 mg mL$^{-1}$ were sonicated during 0.5, 1.0, 2.0, 5.0 and 8.0 hours. After centrifugation at 500 rpm, the graphene concentration $C$ was determined from the measured absorbance by using the Lambert-Beer law,\cite{Coleman2013} $A/l = \alpha C$ with $\alpha = 3620$ $\mathrm{L.g}^{-1}\mathrm{.m}^{-1}$ and $l$ the cell length. After dilution, dispersions appeared darker in color at longer sonication times indicating a higher concentration (see Fig. \ref{fig1}-a). Fig. \ref{fig1}-b shows the evolution of the resulting graphene concentration $C$ as a function of the sonication time $t$.

We find that the data in Fig. \ref{fig1} are well described by the empirical law $C = k t^{1/2}$, where $C$ is the graphene concentration, $t$ is the sonication time (in hours).  The constant $k$ is determined by fitting the above expression. The obtained value is $k = (0.66 \pm 0.03)$ $\mathrm{mg.mL}^{-1}\mathrm{.h}^{-1/2}$, suggesting that the concentration is controlled by the flake size, in good agreement with previously reported values.\cite{Khan2010,Khan2011,Coleman2013} SEM images show the presence of exfoliated flakes (see Fig. \ref{fig1}-c-d). The typical lateral dimension determined from STEM measurements is $800$ nm. The size distribution spans from $200$ nm to $2$ $\mu$m with 70 per cent of the samples below $1$ $\mu$m (see also the supporting information).

\subsection{Raman characterization}

Raman spectroscopy is one of the most useful tools for the characterization of graphene-based materials.\cite{Ferrari2013,Jorio2011} Here we take advantage of this powerful non-destructive tool to characterize our samples. After depositing them onto $SiO_2/Si$ wafers the corresponding Raman spectra were measured at room temperature. Figure \ref{fig2} shows typical results for different samples corresponding to sonication times of 30, 90, 120 and 300 minutes. For reference, the spectrum for the pristine graphite powder is also shown at the bottom. The 2D peak ($\sim  2709$ $\mathrm{cm}^{-1}$) in the spectra of Figure \ref{fig2} can be identified as the typical signal arising from multilayer graphene.\cite{Ferrari2013,Cong2011} The D and G bands were well resolved for all the samples that we measured appearing at$\sim 1353$ $\mathrm{cm}^{-1}$ and  $\sim  1582$ $\mathrm{cm}^{-1}$ respectively.

\begin{figure}[tb]
\includegraphics[width=0.95\columnwidth]{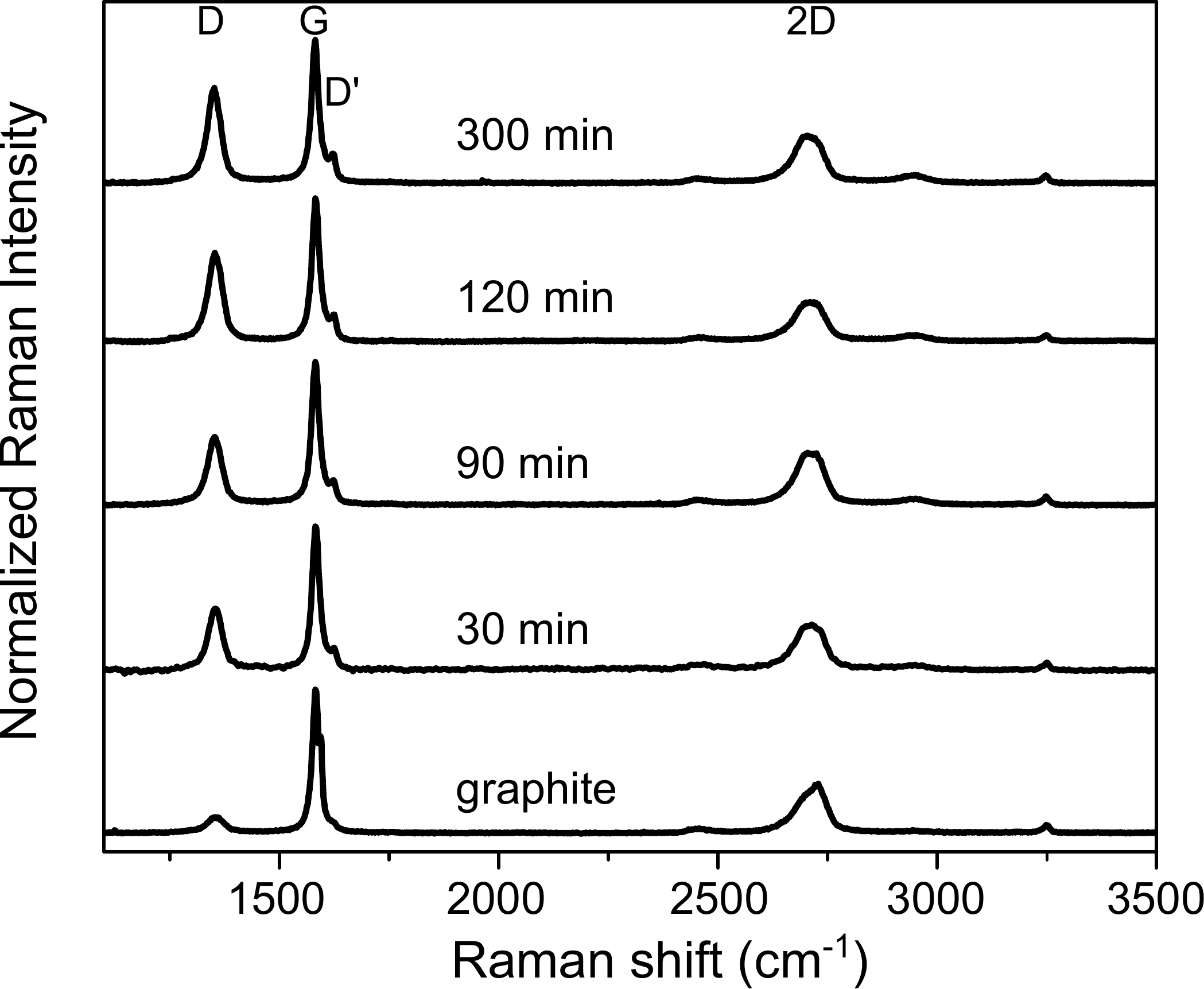}
\caption{Raman spectra of the graphene samples obtained for different sonication times (30, 90, 120 and 300 minutes). The spectrum of a pristine graphite sample is also shown for reference (bottom). All the samples were centrifuged at 500 rpm for 30 min. The spectra show four main signals: D at $\sim 1353$ $\mathrm{cm}^{-1}$, G at $\sim  1582$ $\mathrm{cm}^{-1}$, D' at $\sim  1622$ $\mathrm{cm}^{-1}$ and the 2D band at $\sim  2709$  $\mathrm{cm}^{-1}$. Furthermore, a few combinations of them are also observed: D + D'' at $\sim 2450$ 
$\mathrm{cm}^{-1}$, D + D' at $\sim  2952$ $\mathrm{cm}^{-1}$ and G* at $\sim 3246$ $\mathrm{cm}^{-1}$.}
\label{fig2}
\end{figure}

We emphasize that the Raman spectra show no evidence of graphene oxide formation, as expected for the liquid-phase technique used here. This was additionally tested by EDS spectra measurements. Figure 1-e shows the spectrum corresponding to a 300 min exfoliated graphene sample where the main carbon signal (74.9 wt\%) is observed. Other signals, like oxygen and nitrogen, associated to the retained solvent or other oxygen containing groups, are hardly detected while the Si peaks certainly arise from the substrate.

Defects in the graphene structure break symmetries,\cite{Terrones2010,Book} thereby allowing otherwise forbidden inter/intra-valley processes which lead to the D and D' bands. These defects can be either the sample edges \cite{Eckmann2012,Eckmann2013} or bulk defects \cite{Ferrari2013} and a very important question is whether one could distinguish and quantify different defects based on the Raman spectra \cite{Ferrari2000,Cancado2011,Eckmann2012,Eckmann2013,Ferrari2013}.

The quantity of defects has been shown to be related to the ratio between the D and G bands, $i(D)/i(G)$; the larger the ratio the larger the defect density (and therefore the typical distance between defects).\cite{Ferrari2000,Cancado2011} As we will see below, we observe that $i(D)/i(G)$ increases with increasing the sonication time, thereby indicating an increase in the defect density. The typical distance between defects can be estimated to be in the range between 10 and 20 nm. In the following we focus on a more subtle point concerning the type and location of the emerging defects.

We would like to establish whereas the D and D' signals mainly stem from sample edges and whether this holds for all sonication times $t$. Inspection of the individual Raman signals or of average peak positions/widths for the different values of $t$ did not show any clear trend. Therefore we decided to check for correlations between the amount of disorder as quantified by $i(D)/i(G)$ and the full width at half maximum (FWHM) of the G-band. As we comment below, this correlation provides valuable information on the origin of the D band. The results are shown in Fig. 3 for $t=30$ min (a), $t=90$ min (b), $t=120$ min (c) and $t=300$ min (d). While for short sonication times, the data shows no statistical correlation between $i(D)/i(G)$ and FWHM(G), this changes dramatically in panels (c) and (d) ($t=120, 300$ minutes): as the sonication time increases confidence bands (dashed lines) stretch closer to the best fit (full line) which shows that larger $i(D)/i(G)$ values are correlated with larger widths of the G-band.

\begin{figure}[tb]
\includegraphics[width=0.95\columnwidth]{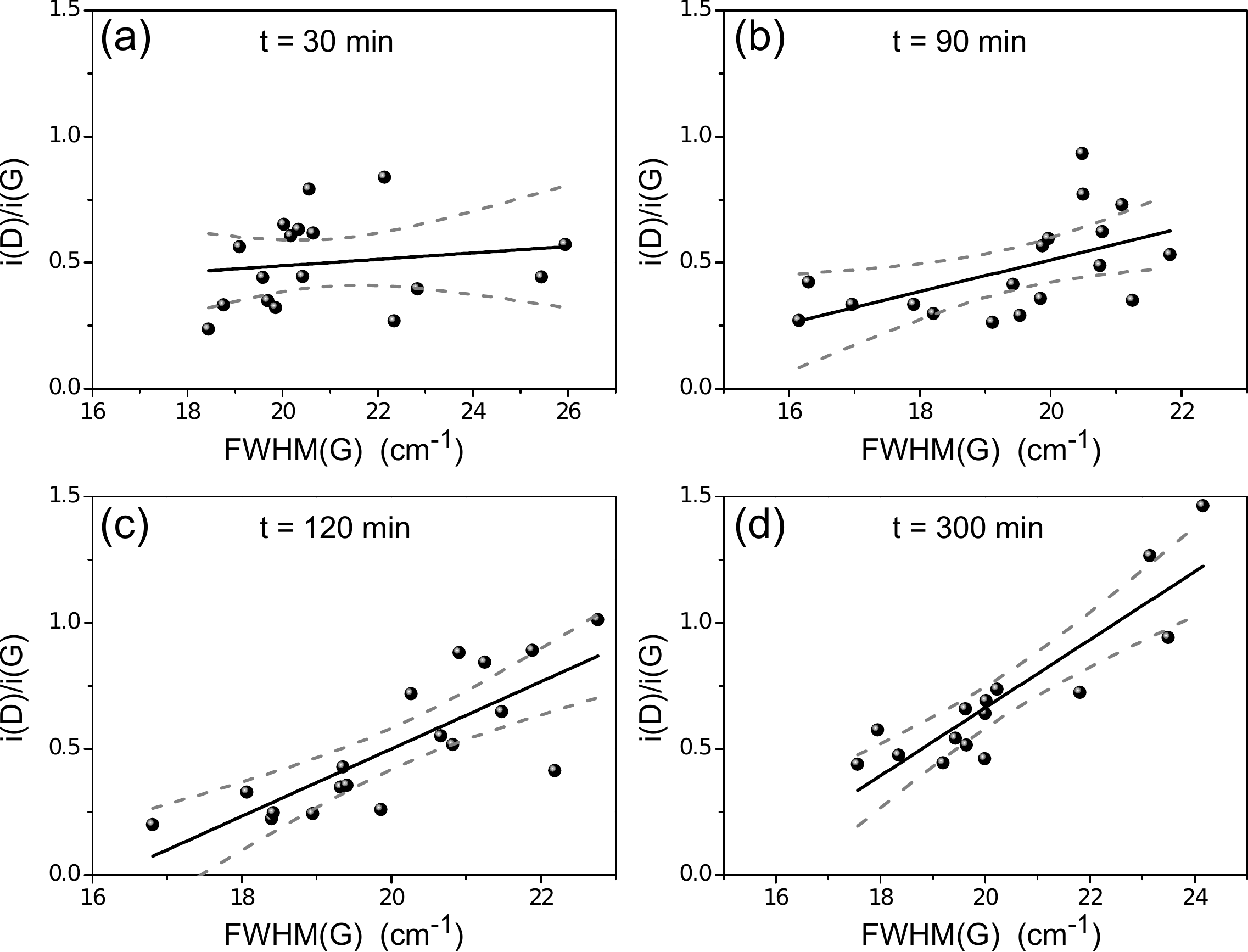}
\caption{$i(D)/i(G)$ versus FWHM(G) plots for dispersions sonicated $30$ (a), $90$ (b), $120$ (c) and $300$ (d) minutes. The dots are experimental data. To assure homogeneity in the conditions, the data in each plot correspond to flakes from the same dispersion. The solid lines are a linear fit to the data, $95$\% confidence bands are also indicated with dashed lines.}
\label{fig3}
\end{figure}

To rationalize the origin and meaning of the observed correlation between  $i(D)/i(G)$ and the width of the G-band we argue on a few key facts. The first one is that a larger $i(D)/i(G)$ ratio indicates a larger amount of either bulk disorder \cite{Ferrari2000,Cancado2011} or edges,\cite{Cancado2004} but it cannot discriminate between them. On the other hand, it turns out that the width of the G band (FWHM(G)) increases with bulk disorder (see for example Fig. 5 of Ref. \onlinecite{Cancado2011} where FWHM(G) is shown to increase when the inter-defect distance decreases) but it does not increase when introducing edges.\cite{Casiraghi2000} Therefore, samples with a larger amount of bulk disorder (exhibiting a larger $i(D)/i(G)$) should also have a larger FWHM(G), \textit{i.e.} $i(D)/i(G)$ should be positively correlated with FWHM(G). In contrast, samples with no bulk disorder should show no correlation between the two magnitudes as reported in Ref. \onlinecite{Torrisi2012}.

These two facts allow us to conclude that the transition observed in our data is likely to be produced by an increase in \textit{bulk disorder} when passing from $t=30$ min to $t=120$ min. These trends are also found if the area of the D and G peaks ($A(D)$ and $A(G)$) are considered, \textit{i.e.}. A(D)/A(G) directly  correlates with FWHM(G). This is shown in Fig. S2 of the supporting information.

\begin{figure}[tb]
\includegraphics[width=0.95\columnwidth]{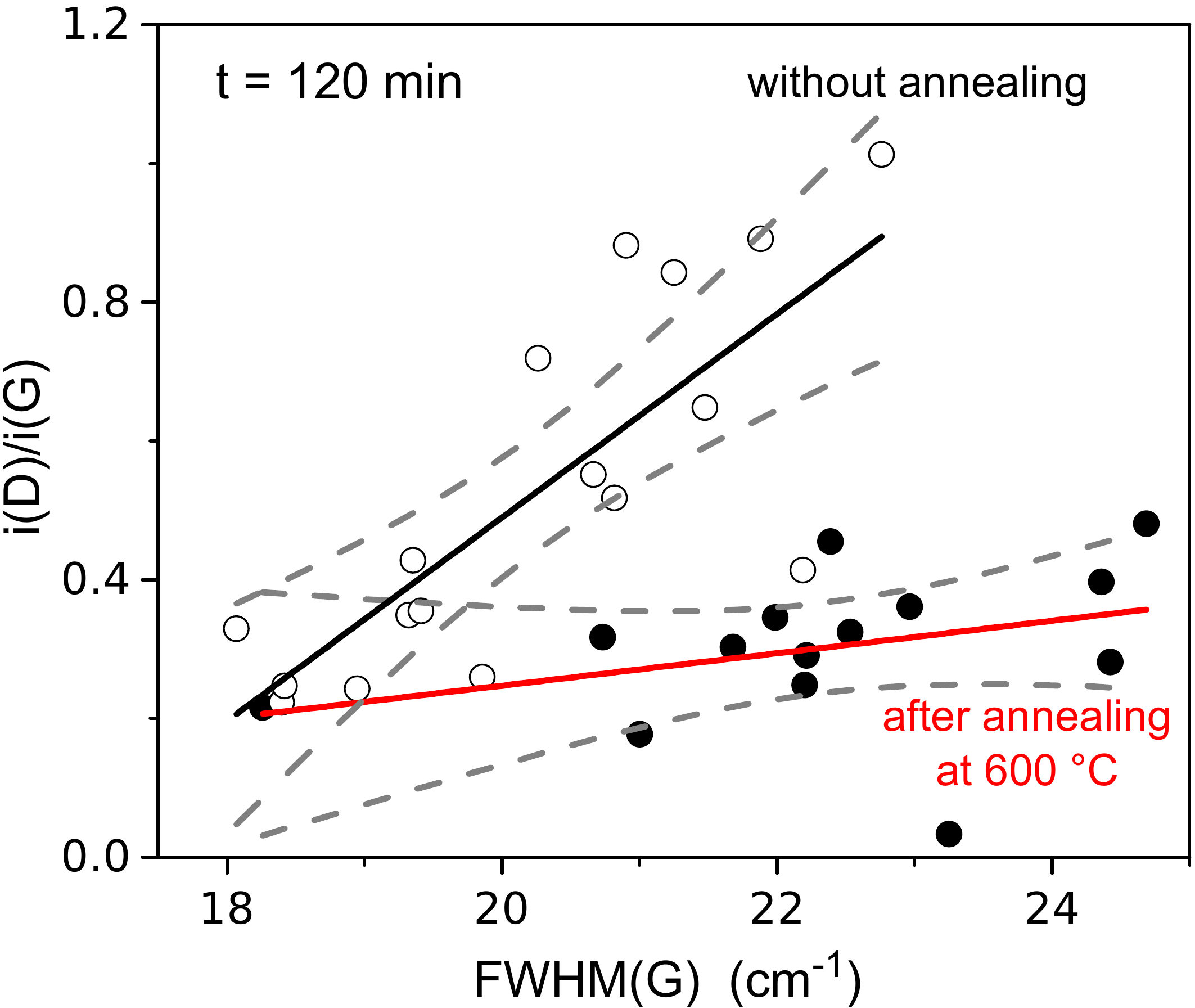}
\caption{$i(D)/i(G)$ as a function of FWHM(G) for a dispersion sonicated for 120 minutes. Empty circles correspond to the samples before annealing while full black circles are for the samples after annealing at $600^{\circ}{\rm C}$. The solid line indicates the best linear fit to the data, $95$\% confidence bands are also indicated in the figure with dashed lines.}
\label{fig4}
\end{figure}

To further confirm the above conclusions we devised a simple additional experiment. If bulk disorder is created for longer sonication times, then annealing the samples should restore a low correlation between $i(D)/i(G)$ versus FWHM(G). Fig. \ref{fig4} shows the results for samples sonicated for $120$ minutes before and after annealing at $600^{\circ}{\rm C}$ for one hour in high vacuum. The data reveals a reduction in the $i(D)/i(G)$ values after annealing, indicating a lower disorder comparable with the values achieved with shorter sonication times (30 and 90 min in Fig. 3 a-b). Whether this smaller $i(D)/i(G)$ ratio is mainly due to bulk defects or not can be inferred from the correlation between $i(D)/i(G)$ and FWHM(G), Fig. 4 shows no correlation between these magnitudes which, as argued before, can only be explained by edges rather than bulk defects. This indicates that most of the bulk defects have been repaired by the annealing process. In contrast, the same thermal treatment applied to flakes sonicated for $30$ minutes does not appreciably change the slope of the relation observed in Fig.3-a (see supplementary information, Fig. S3). In summary, our observations in Figs. 3 and 4 indicate a transition from samples with edge-dominated D band to more disordered structures where bulk defects dominate, the latter emerging when the sonication time is increased.

Up to now, we have presented evidence that in our liquid phase exfoliated samples, bulk defects (in contrast to edges) become dominant (as evidenced in the Raman D band) when the sonication time is of about 120 minutes. A natural question then concerns the specific type of defects produced at this transition. Here we give some possible directions supported by our data. The possible defects include topological defects (like pentagon-heptagon pairs), vacancies, substitutional impurities and sp$^3$-like defects.\cite{Book} Recent studies show that ratio between the D and D' lines is very sensitive to the type of defect \cite{Venezuela2011} with studies reporting a ratio of $3.5$ for boundaries, $7$ for vacancies, $13$ for sp$^3$ and values in-between those for vacancies and sp$^3$ for substitutional impurities.\cite{Eckmann2012,Eckmann2013} The fact that for all sonication times our samples show a roughly constant $i(D)/i(D')$ ratio of ($4.5 \pm 0.5$) rules out vacancies, substitutions and sp$^3$ defects. Moreover, since topological defects have the lowest formation energy \cite{Li2005}, we conclude that this is the most likely defect that emerges as the sonication time is increased. In this sense, different results when using bath sonicator as in our work or a tip sonicator, as reported in Ref. \onlinecite{Torrisi2012}, can not be ruled out.\cite{Buzaglo2013} In any case, lower ultrasound power may help getting better samples.

Based on the conclusions of the previous paragraph, ruling out sp$^3$ or substitutional impurities, a possible mechanism for the creation of bulk defects could be attributed to the cavitation process. Increasing sonication time would then increase the probability of defect formation and therefore the defect density. Numerical simulations may shed light on this issue.

\section{Final remarks}

We present a statistical study of the Raman spectra of graphene multilayers dispersed in NMP by liquid-phase exfoliation. Our results reveal the building-up of bulk-disorder as the sonication time increases. This is reflected in the evolution of the correlation between the ratio of the D to G band intensities ($i(D)/i(G)$) and the width of the G-band. Our results suggest that low disorder samples require a careful tuning of the ultrasonic times. Otherwise, sample annealing may largely enhance the sample's crystalline quality leading to a better material for applications such as composites and conductive inks.

Further analysis of the obtained Raman spectra suggests that the bulk defects are not vacancies, nor substitutional impurities or sp$^3$-like but rather topological defects. The precise mechanism leading to defect formation, which is likely to result from the cavitation process, remains as an interesting subject of study.

\section{Acknowlegdments} 
We acknowledge financial support from ANPCyT Project PICT PRH 61 and SeCyT-UNC. LEFFT acknowledges the support of Trieste's ICTP. M. V. Bracamonte thanks CONICET for the fellowship. MVB thanks to Dr. Arlene O'Neill for useful technical comments. GIL acknowledges projects PICT-324 and PME (2006) 1544. We thank Dr. Paula Bercoff for taking the SEM images presented here as well as Dr. Luis Fabietti for technical help.\\

\section{Supporting information description}
The lateral size distribution of our samples as determines by STEM is included in the supplementary information. $A(D)/A(G)$ versus FWHM(G) is shown in Fig. S2. The annealing experiment for the samples sonicated during 30 minutes is shown in Fig. S3. This material is available free of charge via the Internet at http://pubs.acs.org.


\begin{mcitethebibliography}{30}
\providecommand*\natexlab[1]{#1}
\providecommand*\mciteSetBstSublistMode[1]{}
\providecommand*\mciteSetBstMaxWidthForm[2]{}
\providecommand*\mciteBstWouldAddEndPuncttrue
  {\def\EndOfBibitem{\unskip.}}
\providecommand*\mciteBstWouldAddEndPunctfalse
  {\let\EndOfBibitem\relax}
\providecommand*\mciteSetBstMidEndSepPunct[3]{}
\providecommand*\mciteSetBstSublistLabelBeginEnd[3]{}
\providecommand*\EndOfBibitem{}
\mciteSetBstSublistMode{f}
\mciteSetBstMaxWidthForm{subitem}{(\alph{mcitesubitemcount})}
\mciteSetBstSublistLabelBeginEnd
  {\mcitemaxwidthsubitemform\space}
  {\relax}
  {\relax}

\bibitem[Novoselov et~al.(2004)Novoselov, Geim, Morozov, Jiang, Zhang, Dubonos,
  Grigorieva, and Firsov]{Novoselov2004}
Novoselov,~K.~S.; Geim,~A.~K.; Morozov,~S.~V.; Jiang,~D.; Zhang,~Y.;
  Dubonos,~S.~V.; Grigorieva,~I.~V.; Firsov,~A.~A. Electric Field Effect in
  Atomically Thin Carbon Films. \emph{Science} \textbf{2004}, \emph{306},
  666--669\relax
\mciteBstWouldAddEndPuncttrue
\mciteSetBstMidEndSepPunct{\mcitedefaultmidpunct}
{\mcitedefaultendpunct}{\mcitedefaultseppunct}\relax
\EndOfBibitem
\bibitem[Novoselov et~al.(2005)Novoselov, Geim, Morozov, Jiang, Katsnelson,
  Grigorieva, Dubonos, and Firsov]{Novoselov2005}
Novoselov,~K.~S.; Geim,~A.~K.; Morozov,~S.~V.; Jiang,~D.; Katsnelson,~M.~I.;
  Grigorieva,~I.~V.; Dubonos,~S.~V.; Firsov,~A.~A. Two-dimensional Gas of
  Massless Dirac Fermions in Graphene. \emph{Nature} \textbf{2005}, \emph{438},
  197--200\relax
\mciteBstWouldAddEndPuncttrue
\mciteSetBstMidEndSepPunct{\mcitedefaultmidpunct}
{\mcitedefaultendpunct}{\mcitedefaultseppunct}\relax
\EndOfBibitem
\bibitem[Zhang et~al.(2005)Zhang, Tan, Stormer, and Kim]{Zhang2005}
Zhang,~Y.; Tan,~Y.-W.; Stormer,~H.~L.; Kim,~P. Experimental Observation of the
  Quantum Hall Effect and Berry's Phase in Graphene. \emph{Nature}
  \textbf{2005}, \emph{438}, 201--204\relax
\mciteBstWouldAddEndPuncttrue
\mciteSetBstMidEndSepPunct{\mcitedefaultmidpunct}
{\mcitedefaultendpunct}{\mcitedefaultseppunct}\relax
\EndOfBibitem
\bibitem[Geim and Novoselov(2007)Geim, and Novoselov]{Geim2007}
Geim,~A.~K.; Novoselov,~K.~S. The Rise of Graphene. \emph{Nature Materials}
  \textbf{2007}, \emph{6}, 183--191\relax
\mciteBstWouldAddEndPuncttrue
\mciteSetBstMidEndSepPunct{\mcitedefaultmidpunct}
{\mcitedefaultendpunct}{\mcitedefaultseppunct}\relax
\EndOfBibitem
\bibitem[Balandin et~al.(2008)Balandin, Ghosh, Bao, Calizo, Teweldebrhan, Miao,
  and Lau]{Balandin2008}
Balandin,~A.~A.; Ghosh,~S.; Bao,~W.; Calizo,~I.; Teweldebrhan,~D.; Miao,~F.;
  Lau,~C.~N. Superior Thermal Conductivity of Single-Layer Graphene. \emph{Nano
  Lett.} \textbf{2008}, \emph{8}, 902--907\relax
\mciteBstWouldAddEndPuncttrue
\mciteSetBstMidEndSepPunct{\mcitedefaultmidpunct}
{\mcitedefaultendpunct}{\mcitedefaultseppunct}\relax
\EndOfBibitem
\bibitem[Lee et~al.(2008)Lee, Wei, Kysar, and Hone]{Lee2008}
Lee,~C.; Wei,~X.; Kysar,~J.~W.; Hone,~J. Measurement of the Elastic Properties
  and Intrinsic Strength of Monolayer Graphene. \emph{Science} \textbf{2008},
  \emph{321}, 385--388\relax
\mciteBstWouldAddEndPuncttrue
\mciteSetBstMidEndSepPunct{\mcitedefaultmidpunct}
{\mcitedefaultendpunct}{\mcitedefaultseppunct}\relax
\EndOfBibitem
\bibitem[Foa~Torres et~al.(2014)Foa~Torres, Roche, and Charlier]{Book}
Foa~Torres,~L. E.~F.; Roche,~S.; Charlier,~J.~C. \emph{Introduction to
  Graphene-Based Nanomaterials: From Electronic Structure to Quantum
  Transport}; Cambridge University Press, 2014\relax
\mciteBstWouldAddEndPuncttrue
\mciteSetBstMidEndSepPunct{\mcitedefaultmidpunct}
{\mcitedefaultendpunct}{\mcitedefaultseppunct}\relax
\EndOfBibitem
\bibitem[Novoselov et~al.(2012)Novoselov, Falko, Colombo, Gellert, Schwab, and
  Kim]{Novoselov2012}
Novoselov,~K.~S.; Falko,~V.~I.; Colombo,~L.; Gellert,~P.~R.; Schwab,~M.~G.;
  Kim,~K. A Roadmap for Graphene. \emph{Nature} \textbf{2012}, \emph{490},
  192--200\relax
\mciteBstWouldAddEndPuncttrue
\mciteSetBstMidEndSepPunct{\mcitedefaultmidpunct}
{\mcitedefaultendpunct}{\mcitedefaultseppunct}\relax
\EndOfBibitem
\bibitem[Bae et~al.(2010)Bae, Kim, Lee, Xu, Park, Zheng, Balakrishnan, Lei,
  Ri~Kim, Song, Kim, Kim, Ozyilmaz, Ahn, Hong, and Iijima]{Bae2010}
Bae,~S.; Kim,~H.; Lee,~Y.; Xu,~X.; Park,~J.-S.; Zheng,~Y.; Balakrishnan,~J.; Lei,~T.; Ri~Kim,~H.; Song,~Y.~I. \textit{et~al.} Roll-to-roll Production of 30-inch Graphene Films for
  Transparent Electrodes. \emph{Nature Nanotechnology} \textbf{2010}, \emph{5},
  574--578\relax
\mciteBstWouldAddEndPuncttrue
\mciteSetBstMidEndSepPunct{\mcitedefaultmidpunct}
{\mcitedefaultendpunct}{\mcitedefaultseppunct}\relax
\EndOfBibitem
\bibitem[Emtsev et~al.(2009)Emtsev, Bostwick, Horn, Jobst, Kellogg, Ley,
  McChesney, Ohta, Reshanov, Rohrl, Rotenberg, Schmid, Waldmann, Weber, and
  Seyller]{Emtsev2009}
Emtsev,~K.~V.; Bostwick,~A.; Horn,~K.; Jobst,~J.; Kellogg,~G.~L.; Ley,~L.;
  McChesney,~J.~L.; Ohta,~T.; Reshanov,~S.~A.; Rohrl,~J. \textit{et al.}  Towards Wafer-size
  Graphene Layers by Atmospheric Pressure Graphitization of Silicon Carbide.
  \emph{Nature Materials} \textbf{2009}, \emph{8}, 203--207\relax
\mciteBstWouldAddEndPuncttrue
\mciteSetBstMidEndSepPunct{\mcitedefaultmidpunct}
{\mcitedefaultendpunct}{\mcitedefaultseppunct}\relax
\EndOfBibitem
\bibitem[Torrisi et~al.(2012)Torrisi, Hasan, Wu, Sun, Lombardo, Kulmala, Hsieh,
  Jung, Bonaccorso, Paul, Chu, and Ferrari]{Torrisi2012}
Torrisi,~F.; Hasan,~T.; Wu,~W.; Sun,~Z.; Lombardo,~A.; Kulmala,~T.~S.;
  Hsieh,~G.-W.; Jung,~S.; Bonaccorso,~F.; Paul,~P.~J. \textit{et~al.}
  Inkjet-Printed Graphene Electronics. \emph{ACS Nano} \textbf{2012}, \emph{6},
  2992--3006\relax
\mciteBstWouldAddEndPuncttrue
\mciteSetBstMidEndSepPunct{\mcitedefaultmidpunct}
{\mcitedefaultendpunct}{\mcitedefaultseppunct}\relax
\EndOfBibitem
\bibitem[Han et~al.(2013)Han, Chen, Zhu, Preston, Wan, Fang, and Hu]{Han2013}
Han,~X.; Chen,~Y.; Zhu,~H.; Preston,~C.; Wan,~J.; Fang,~Z.; Hu,~L. Scalable,
  Printable, Surfactant-free Graphene Ink Directly from Graphite.
  \emph{Nanotechnology} \textbf{2013}, \emph{24}, 205304--\relax
\mciteBstWouldAddEndPuncttrue
\mciteSetBstMidEndSepPunct{\mcitedefaultmidpunct}
{\mcitedefaultendpunct}{\mcitedefaultseppunct}\relax
\EndOfBibitem
\bibitem[Nicolosi et~al.(2013)Nicolosi, Chhowalla, Kanatzidis, Strano, and
  Coleman]{Nicolosi2013}
Nicolosi,~V.; Chhowalla,~M.; Kanatzidis,~M.~G.; Strano,~M.~S.; Coleman,~J.~N.
  Liquid Exfoliation of Layered Materials. \emph{Science} \textbf{2013},
  \emph{340}, --\relax
\mciteBstWouldAddEndPuncttrue
\mciteSetBstMidEndSepPunct{\mcitedefaultmidpunct}
{\mcitedefaultendpunct}{\mcitedefaultseppunct}\relax
\EndOfBibitem
\bibitem[Khan et~al.(2011)Khan, Porwal, O~Neill, Nawaz, May, and
  Coleman]{Khan2011}
Khan,~U.; Porwal,~H.; O~Neill,~A.; Nawaz,~K.; May,~P.; Coleman,~J.~N.
  Solvent-Exfoliated Graphene at Extremely High Concentration. \emph{Langmuir}
  \textbf{2011}, \emph{27}, 9077--9082\relax
\mciteBstWouldAddEndPuncttrue
\mciteSetBstMidEndSepPunct{\mcitedefaultmidpunct}
{\mcitedefaultendpunct}{\mcitedefaultseppunct}\relax
\EndOfBibitem
\bibitem[Paton et~al.(2014)Paton, Varrla, Backes, Smith, Khan, O'Neill, Boland,
  Lotya, Istrate, King, Higgins, Barwich, May, Puczkarski, Ahmed, Moebius,
  Pettersson, Long, Coelho, O'Brien, McGuire, Sanchez, Duesberg, McEvoy,
  Pennycook, Downing, Crossley, Nicolosi, and Coleman]{Paton2014}
Paton,~K.~R.; Varrla,~E.; Backes,~C.; Smith,~R.~J.; Khan,~U.; O'Neill,~A.; Boland,~C.; Lotya,~M.;  Istrate,~O.~M.; King,~P. \textit{et~al.}  Scalable Production of Large Quantities of Defect-free
  Few-layer Graphene by Shear Exfoliation in Liquids. \emph{Nature Materials}
  \textbf{2014}, \emph{13}, 624--630\relax
\mciteBstWouldAddEndPuncttrue
\mciteSetBstMidEndSepPunct{\mcitedefaultmidpunct}
{\mcitedefaultendpunct}{\mcitedefaultseppunct}\relax
\EndOfBibitem
\bibitem[Stankovich et~al.(2006)Stankovich, Dikin, Dommett, Kohlhaas, Zimney,
  Stach, Piner, Nguyen, and Ruoff]{Stankovich2006}
Stankovich,~S.; Dikin,~D.~A.; Dommett,~G. H.~B.; Kohlhaas,~K.~M.;
  Zimney,~E.~J.; Stach,~E.~A.; Piner,~R.~D.; Nguyen,~S.~T.; Ruoff,~R.~S.
  Graphene-based Composite Materials. \emph{Nature} \textbf{2006}, \emph{442},
  282--286\relax
\mciteBstWouldAddEndPuncttrue
\mciteSetBstMidEndSepPunct{\mcitedefaultmidpunct}
{\mcitedefaultendpunct}{\mcitedefaultseppunct}\relax
\EndOfBibitem
\bibitem[Khan et~al.(2010)Khan, O~Neill, Lotya, De, and Coleman]{Khan2010}
Khan,~U.; O~Neill,~A.; Lotya,~M.; De,~S.; Coleman,~J.~N. High-Concentration
  Solvent Exfoliation of Graphene. \emph{Small} \textbf{2010}, \emph{6},
  864--871\relax
\mciteBstWouldAddEndPuncttrue
\mciteSetBstMidEndSepPunct{\mcitedefaultmidpunct}
{\mcitedefaultendpunct}{\mcitedefaultseppunct}\relax
\EndOfBibitem
\bibitem[Coleman(2013)]{Coleman2013}
Coleman,~J.~N. Liquid Exfoliation of Defect-Free Graphene. \emph{Acc. Chem.
  Res.} \textbf{2013}, \emph{46}, 14--22\relax
\mciteBstWouldAddEndPuncttrue
\mciteSetBstMidEndSepPunct{\mcitedefaultmidpunct}
{\mcitedefaultendpunct}{\mcitedefaultseppunct}\relax
\EndOfBibitem
\bibitem[Ferrari and Basko(2013)Ferrari, and Basko]{Ferrari2013}
Ferrari,~A.~C.; Basko,~D.~M. Raman Spectroscopy as a Versatile Tool for
  Studying the Properties of Graphene. \emph{Nature Nanotechnology} \textbf{2013}, \emph{8},
  235--246\relax
\mciteBstWouldAddEndPuncttrue
\mciteSetBstMidEndSepPunct{\mcitedefaultmidpunct}
{\mcitedefaultendpunct}{\mcitedefaultseppunct}\relax
\EndOfBibitem
\bibitem[Jorio et~al.(2011)Jorio, Dressehlhaus, Saito, and
  Dresselhaus]{Jorio2011}
Jorio,~A.; Dressehlhaus,~M.; Saito,~R.; Dresselhaus,~G.~F. \emph{Raman
  Spectroscopy in Graphene Related Systems}; Wiley-VCH, 2011\relax
\mciteBstWouldAddEndPuncttrue
\mciteSetBstMidEndSepPunct{\mcitedefaultmidpunct}
{\mcitedefaultendpunct}{\mcitedefaultseppunct}\relax
\EndOfBibitem
\bibitem[Cong et~al.(2011)Cong, Yu, Saito, Dresselhaus, and
  Dresselhaus]{Cong2011}
Cong,~C.; Yu,~T.; Saito,~R.; Dresselhaus,~G.~F.; Dresselhaus,~M.~S.
  Second-Order Overtone and Combination Raman Modes of Graphene Layers in the
  Range of 1690-2150 cm$^-1$. \emph{ACS Nano} \textbf{2011}, \emph{5},
  1600--1605\relax
\mciteBstWouldAddEndPuncttrue
\mciteSetBstMidEndSepPunct{\mcitedefaultmidpunct}
{\mcitedefaultendpunct}{\mcitedefaultseppunct}\relax
\EndOfBibitem
\bibitem[Terrones et~al.(2010)Terrones, Botello-M\'endez, Campos-Delgado,
  L\'opez-Ur\'{\i}as, Vega-Cant\'u, Rodr\'{\i}guez-Mac\'{\i}as, El\'{\i}as,
  Mu\~noz Sandoval, Cano-Marquez, Charlier, and Terrones]{Terrones2010}
Terrones,~M.; Botello-M\'endez,~A.~R.; Campos-Delgado,~J.;
  L\'opez-Ur\'{\i}as,~F.; Vega-Cant\'u,~Y.~I.;
  Rodr\'{\i}guez-Mac\'{\i}as,~F.~J.; El\'{\i}as,~A.~L.; Mu\~noz Sandoval,~E.;
  Cano-Marquez,~A.~G.; Charlier,~J.-C. \textit{et al.} Graphene and Graphite
  Nanoribbons: Morphology, Properties, Synthesis, Defects and Applications.
  \emph{Nano Today} \textbf{2010}, \emph{5}, 351--372\relax
\mciteBstWouldAddEndPuncttrue
\mciteSetBstMidEndSepPunct{\mcitedefaultmidpunct}
{\mcitedefaultendpunct}{\mcitedefaultseppunct}\relax
\EndOfBibitem
\bibitem[Eckmann et~al.(2012)Eckmann, Felten, Mishchenko, Britnell, Krupke,
  Novoselov, and Casiraghi]{Eckmann2012}
Eckmann,~A.; Felten,~A.; Mishchenko,~A.; Britnell,~L.; Krupke,~R.;
  Novoselov,~K.~S.; Casiraghi,~C. Probing the Nature of Defects in Graphene by
  Raman Spectroscopy. \emph{Nano Letters} \textbf{2012}, \emph{12},
  3925--3930\relax
\mciteBstWouldAddEndPuncttrue
\mciteSetBstMidEndSepPunct{\mcitedefaultmidpunct}
{\mcitedefaultendpunct}{\mcitedefaultseppunct}\relax
\EndOfBibitem
\bibitem[Eckmann et~al.(2013)Eckmann, Felten, Verzhbitskiy, Davey, and
  Casiraghi]{Eckmann2013}
Eckmann,~A.; Felten,~A.; Verzhbitskiy,~I.; Davey,~R.; Casiraghi,~C. Raman Study
  on Defective Graphene: Effect of the Excitation Energy, Type, and Amount of
  Defects. \emph{Phys. Rev. B} \textbf{2013}, \emph{88}, 035426\relax
\mciteBstWouldAddEndPuncttrue
\mciteSetBstMidEndSepPunct{\mcitedefaultmidpunct}
{\mcitedefaultendpunct}{\mcitedefaultseppunct}\relax
\EndOfBibitem
\bibitem[Ferrari and Robertson(2000)Ferrari, and Robertson]{Ferrari2000}
Ferrari,~A.~C.; Robertson,~J. Interpretation of Raman Spectra of Disordered and
  Amorphous Carbon. \emph{Phys. Rev. B} \textbf{2000}, \emph{61},
  14095--14107\relax
\mciteBstWouldAddEndPuncttrue
\mciteSetBstMidEndSepPunct{\mcitedefaultmidpunct}
{\mcitedefaultendpunct}{\mcitedefaultseppunct}\relax
\EndOfBibitem
\bibitem[Cancado et~al.(2011)Cancado, Jorio, Ferreira, Stavale, Achete, Capaz,
  Moutinho, Lombardo, Kulmala, and Ferrari]{Cancado2011}
Cancado,~L.~G.; Jorio,~A.; Ferreira,~E. H.~M.; Stavale,~F.; Achete,~C.~A.;
  Capaz,~R.~B.; Moutinho,~M. V.~O.; Lombardo,~A.; Kulmala,~T.~S.;
  Ferrari,~A.~C. Quantifying Defects in Graphene via Raman Spectroscopy at
  Different Excitation Energies. \emph{Nano Lett.} \textbf{2011}, \emph{11},
  3190--3196\relax
\mciteBstWouldAddEndPuncttrue
\mciteSetBstMidEndSepPunct{\mcitedefaultmidpunct}
{\mcitedefaultendpunct}{\mcitedefaultseppunct}\relax
\EndOfBibitem
\bibitem[Cancado et~al.(2004)Cancado, Pimenta, Neves, Dantas, and
  Jorio]{Cancado2004}
Cancado,~L.~G.; Pimenta,~M.~A.; Neves,~B. R.~A.; Dantas,~M. S.~S.; Jorio,~A.
  Influence of the Atomic Structure on the Raman Spectra of Graphite Edges.
  \emph{Phys. Rev. Lett.} \textbf{2004}, \emph{93}, 247401--\relax
\mciteBstWouldAddEndPuncttrue
\mciteSetBstMidEndSepPunct{\mcitedefaultmidpunct}
{\mcitedefaultendpunct}{\mcitedefaultseppunct}\relax
\EndOfBibitem
\bibitem[Venezuela et~al.(2011)Venezuela, Lazzeri, and Mauri]{Venezuela2011}
Venezuela,~P.; Lazzeri,~M.; Mauri,~F. Theory of Double-resonant Raman Spectra
  in Graphene: Intensity and Line Shape of Defect-induced and Two-phonon Bands.
  \emph{Phys. Rev. B} \textbf{2011}, \emph{84}, 035433--\relax
\mciteBstWouldAddEndPuncttrue
\mciteSetBstMidEndSepPunct{\mcitedefaultmidpunct}
{\mcitedefaultendpunct}{\mcitedefaultseppunct}\relax
\EndOfBibitem
\bibitem[Li et~al.(2005)Li, Reich, and Robertson]{Li2005}
Li,~L.; Reich,~S.; Robertson,~J. Defect Energies of Graphite:
  Density-functional Calculations. \emph{Phys. Rev. B} \textbf{2005},
  \emph{72}, 184109--\relax
\mciteBstWouldAddEndPuncttrue
\mciteSetBstMidEndSepPunct{\mcitedefaultmidpunct}
{\mcitedefaultendpunct}{\mcitedefaultseppunct}\relax
\EndOfBibitem
\bibitem{Casiraghi2000}
Casiraghi,~C.; Hartschuh,~A.; Qian,~H; Piscanec,~C.; Georgi,~C.; Fasoli,~A.; Novoselov,~K.~S.; Basko,~D.~M.; Ferrari,~A.~C. Raman Spectroscopy of Graphene Edges. 
  \emph{NanoLett.} \textbf{2009}, \emph{9}, 1433--1441\relax
\mciteBstWouldAddEndPuncttrue
\mciteSetBstMidEndSepPunct{\mcitedefaultmidpunct}
{\mcitedefaultendpunct}{\mcitedefaultseppunct}\relax
\EndOfBibitem
\bibitem{Buzaglo2013}
Buzaglo,~M.; Shtein,~M.; Kober,~S.; Lovrincic,~R; Vilan,~A; Regev,~O 
Critical Parameters in Exfoliating Graphite into Graphene.
\emph{Phys. Chem. Chem. Phys.} \textbf{2013},
  \emph{15}, 4428--4435\relax
\mciteBstWouldAddEndPuncttrue
\mciteSetBstMidEndSepPunct{\mcitedefaultmidpunct}
{\mcitedefaultendpunct}{\mcitedefaultseppunct}\relax
\EndOfBibitem
\end{mcitethebibliography}

\providecommand*\mcitethebibliography{\thebibliography}
\csname @ifundefined\endcsname{endmcitethebibliography}
  {\let\endmcitethebibliography\endthebibliography}{}

\newpage

\textbf{TOC Image}

\begin{figure}[tbhc]
\includegraphics[width=0.8\columnwidth]{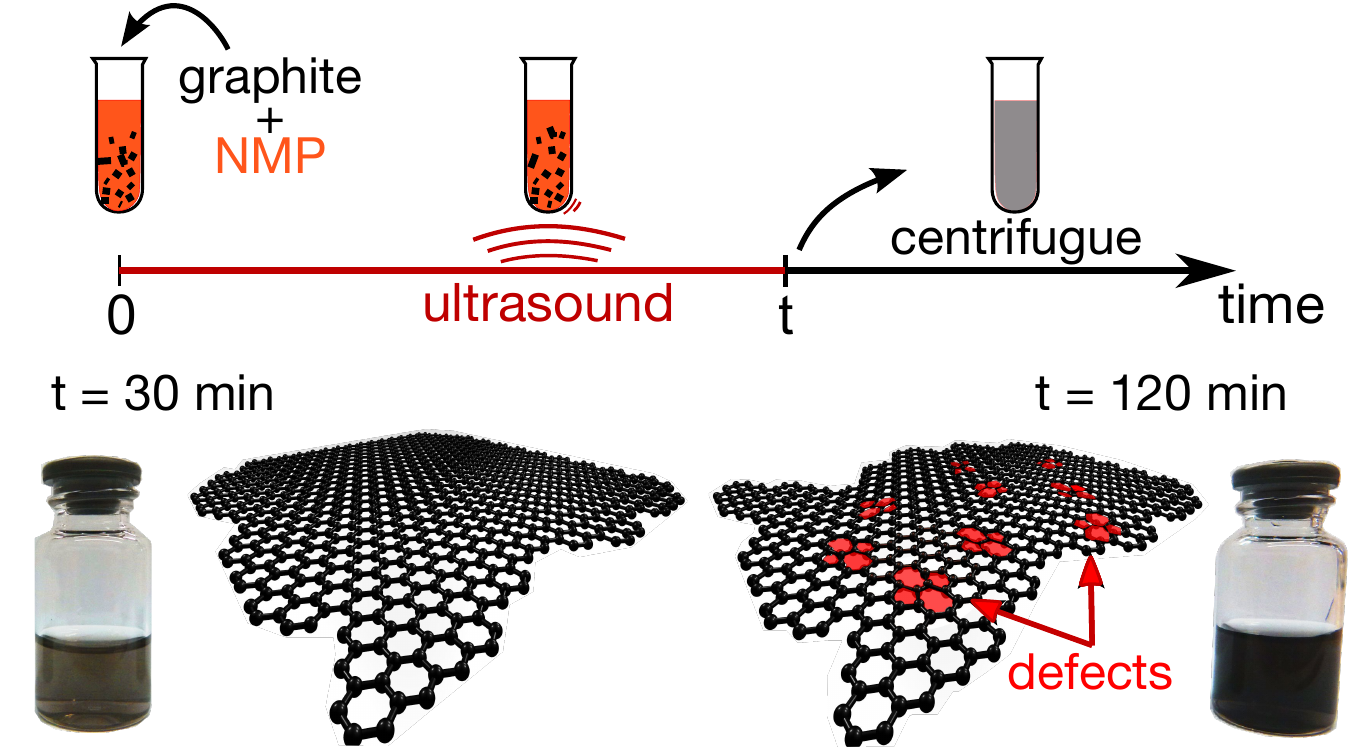}
\label{toc}
\end{figure}
\setcounter{figure}{0}
\renewcommand\thefigure{S\arabic{figure}}    

\section{Supporting Information}
\begin{figure}[tbph]
\includegraphics[width=0.95\columnwidth]{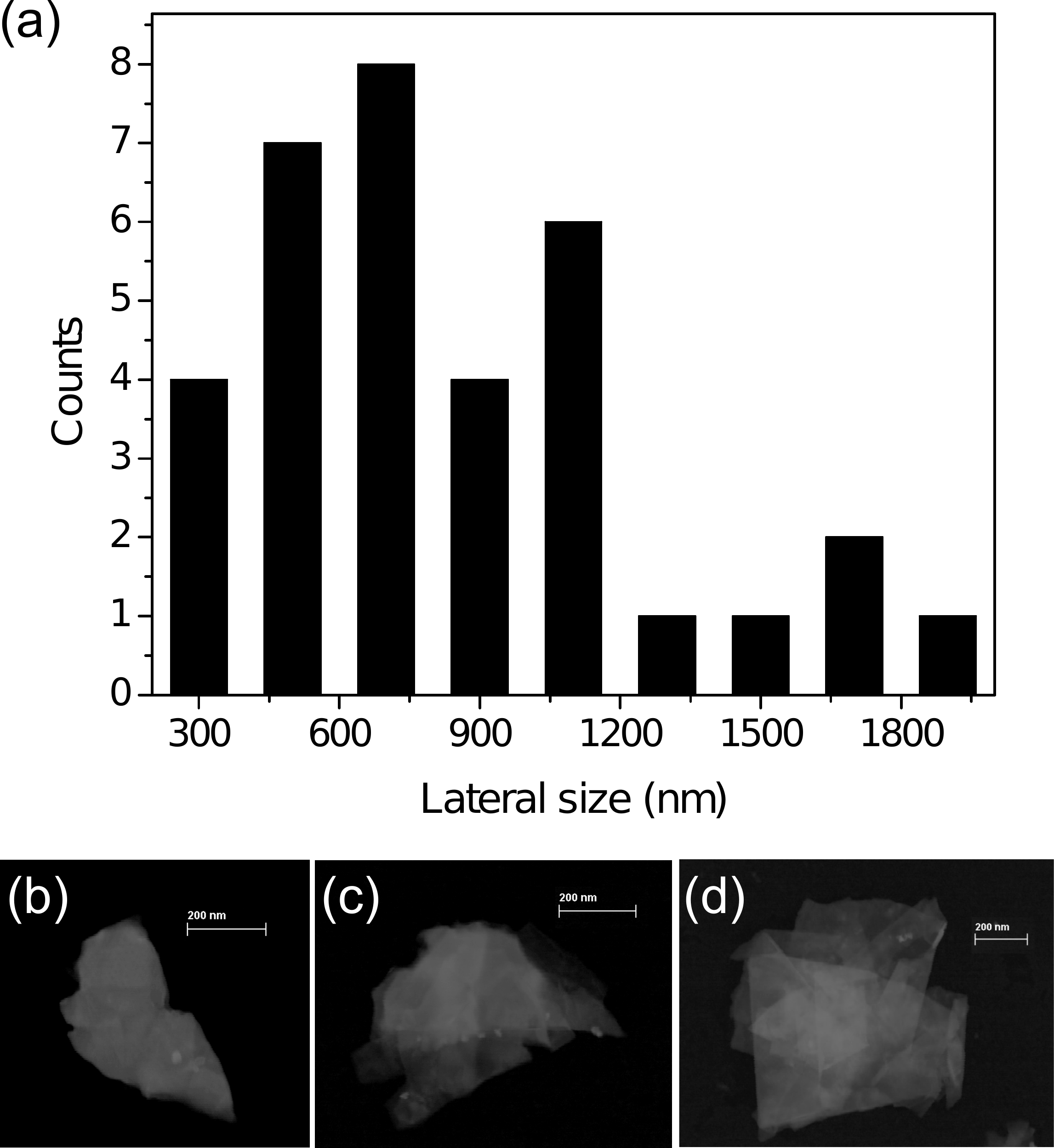}
\caption{(a) Histogram with lateral size distribution for the flakes measured from STEM images, (b-d) images of representative flakes.}
\label{figsup}
\end{figure}

The typical dimensions of the exfoliated flakes were measured using STEM. To do so we deposited a drop (10 $\mu l$) on the surface of a carbon coated copper grid (400mesh) and then captured the STEM images at 25kV using a $\Sigma$igma FE-SEM. Images were processed with ImageJ software. The measured lateral size distribution as well a few representative images are included in Fig. 1. One can see that the size distribution spans from 200 nm to 2 microns with 70 per cent of the samples falling below 1 micron. The white bar in Fig.  S1 b-d has 200 nm.

Figure S2 shows A(D)/A(G) versus FWHM(G) using the same experimental data as in Fig. 3 of the main text. The trend observed in the main text is preserved.

\begin{figure}[tb]
\center
\includegraphics[width=0.9\columnwidth]{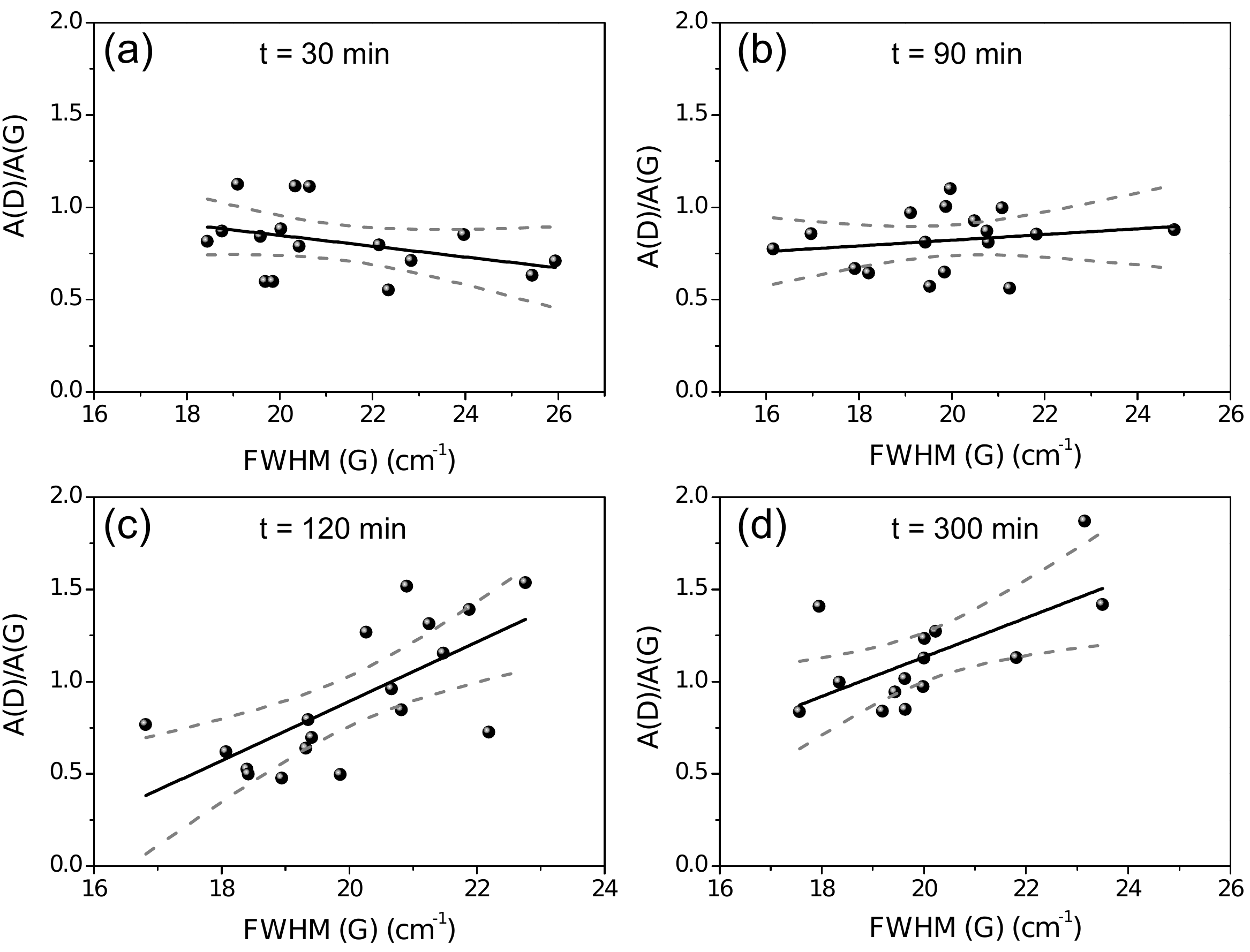}
\caption{$A(D)/A(G)$ versus FWHM(G) plots for dispersions sonicated $30$ (a), $90$ (b), $120$ (c) and $300$ (d) minutes. The dots are experimental data. To assure homogeneity in the conditions, the data in each plot correspond to flakes from the same dispersion. The full lines are a linear fit to the data, $95$\% confidence bands are also indicated with dashed lines. We observe the same trend shown in Fig. 3 of the main text for $i(D)/i(G)$ versus FWHM(G).}
\label{fig2}
\end{figure}

In the main text we discussed the annealing experiment and showed results for 120 minutes of ultrasound. Figure S3 shows $i(D)/i(G)$ as a function of FWHM(G) for dispersion sonicated 30 minutes. We observe that the slope of these curves before and after annealing do not change within the experimental error, as expected from the arguments presented in the main text.

\begin{figure}[tbph]
\center
\includegraphics[width=0.9\columnwidth]{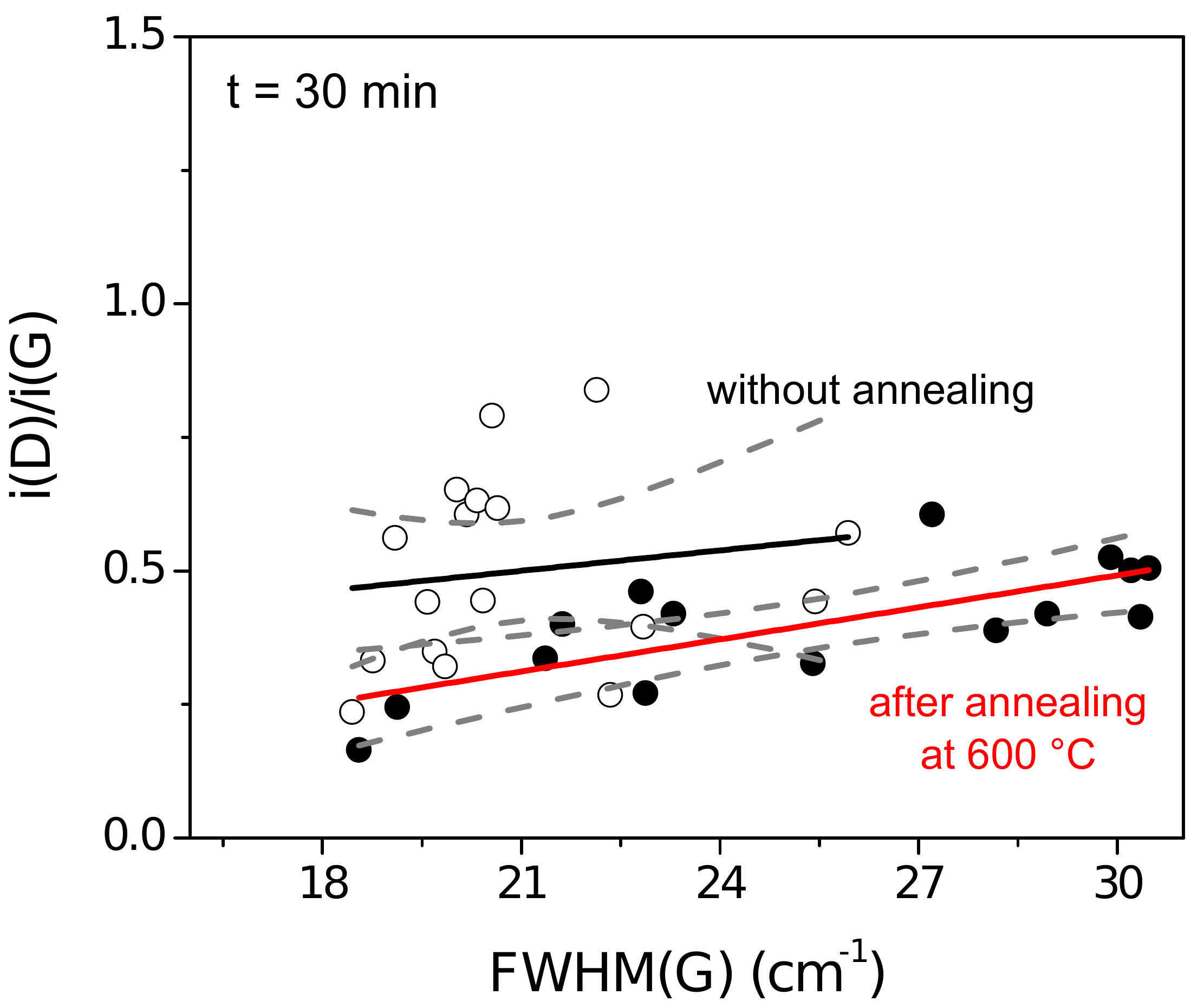}
\caption{$i(D)/i(G)$ as a function of FWHM(G) for dispersion sonicated 30 minutes. Empty circles correspond to the samples before annealing while full black circles are for the samples after annealing at $600^{\circ}{\rm C}$. The full indicates the best linear fit to the data, $95$\% confidence bands are also indicated in the figure with dashed lines.}
\label{fig3}
\end{figure}
\end{document}